
\input amstex
\magnification=1200
\documentstyle{amsppt}
\NoRunningHeads
\font\cyr=wncyr10
\font\cyb=wncyb10
\font\cyi=wncyi10
\font\cyre=wncyr8
\font\cyrs=wncyr6
\font\cyrf=wncyr6 at 5pt
\topmatter
\title\cyb K opisaniyu interaktivnyh psihoinformatsionnyh sistem.
Predvaritelp1nye zamechaniya.
\endtitle
\author\cyr D.V.Yurp1ev
\endauthor
\date adap-org/9409003
\enddate
\address\cyre Otdel matematiki,\newline
Nauchno--issledovatelp1skie0 institut\newline
sistemnyh issledovanie0 RAN, Moskva
\endaddress
\endtopmatter
\document\cyr
Kak otmechalosp1 v rabote avtora [1] iskusstvennye interaktivnye
kom\-pp1yuterograficheskie psihoinformatsionnye sistemy predstavlyayut
in\-te\-res kak yazyk matematicheskoe0 fiziki dlya opisaniya razlichnyh
estestvennyh (v tom chisle sensornyh, vizualp1nyh ili respiratornyh)
interaktivnyh psihoinformatsionnyh sistem. Klyuchevuyu rolp1 v podobnom
opisa\-nii
igraet sozdanie dinamicheskogo izobrazheniya protekayushchego v estest\-vennoe0
psihoinformacionnoe0 sisteme interaktivnogo processa v
in\-ten\-tsi\-o\-nalp1\-noe0
anomalp1noe0 virtualp1noe0 realp1nosti [2],
realizuyushchee0 kom\-pp1yu\-te\-ro\-gra\-fi\-ches\-kie0 interfee0s
nekotoroe0 iskusstvennoe0 interaktivnoe0 sistemy.

V dannoe0 rabote my ogranichivaemsya uzkim klassom estestvennyh interaktivnyh
psihoinformatsionnyh sistem, kotorye dopuskayut vydelenie aktivnoe0 i
passivnoe0 sostavlyayushchee0: aktivnogo agenta -- subp2ekta i passivnogo
agenta -- obp2ekta. Takoe0 klass estestvennyh interaktivnyh
psihoinformatsionnyh sistem naibolee blizok k sistemam, dopuskayushchim odnu iz
klassicheskih form
opisaniya (formulp1nogo ili graficheskogo, staticheskogo ili dinamicheskogo,
algoritmicheskogo, determinirovannogo ili veroyatnostnogo).

\definition{\cyb Opredelenie} \cyr {\cyi Izobrazheniem estestvennoe0
interaktivnoe0 psihoinformatsionnoe sistemy posredstvom iskusstvennoe0
interaktivnoe0 kom\-pp1yu\-te\-ro\-gra\-fi\-che\-skoe0 psihoinformatsionnoe0
sistemy\/} nazyvaet\-sya zadanie algoritma postroeniya dinamicheskogo
izobrazheniya proizvolp1nogo pro\-te\-ka\-yu\-shche\-go v estestvennoe0
sisteme interaktivnogo protsesssa v intentsionalp1noe0 anomalp1noe0
virtualp1noe0 realp1nosti, realizuyushchee0
kom\-pp1yu\-te\-ro\-gra\-fi\-che\-s\-kie0 interfee0s iskusstvennoe0
interaktivnoe0 sistemy.
\enddefinition

\definition{\cyb Opredelenie} \cyr {\cyi Virtualizatsiee0 estestvennoe0
interaktivnoe0 psihoinformatsionnoe0 sistemy\/} nazyvaet\-sya sopostavlenie
ee0 ei0 izobrazheniya posredstvom iskusstvennoe0 interaktivnoe0
kompp1yuterograficheskoe0 psihoinformatsinnoe0 sistemy po opredeli0nnoe0
sovokupnosti e1ksperimentalp1nyh dannyh o sisteme v ustanovivshemsya
avtokolebatelp1nom rezhime.
\enddefinition
Inymi slovami, virtualizatsiya estestvennoe0 interaktivnoe0
psihoinformatsionnoe0 sistemy pozvolyaet po harakteristikam nekotorogo
og\-ra\-ni\-chen\-no\-go mnozhestva avtokolebatelp1nyh interaktivnyh
protsessov for\-mi\-ro\-vatp1 dinamicheskoe izobrazhenie proizvolp1nogo (ne
obyazatelp1no avtokolebatelp1nogo) interaktivnogo protsessa v
intentsionalp1noe0 ano\-malp1\-noe0 virtualp1noe0 realp1nosti,
realizuyushchee0 kompp1yuterograficheskie0 interfee0s nekotoroe0
iskusstvennoe0 interaktivnoe0 psihoinformatsionnoe0 sistemy.

\definition{\cyb Poyasnenie} \cyr Vybor termina "virtualizatsiya" svyazan s
tem, chto dinamicheskoe izobrazhenie interaktivnogo protsessa predstavlyaet
soboe0 ano\-malp1\-nuyu virtualp1nuyu realp1nostp1 v shirokom smysle e1togo
ponyatiya (sm.~[2]).
\enddefinition
Vvidu deleniya interesuyushchih nas estestvennyh interaktivnyh
psihoinformatsionnyh sistem na aktivnuyu i passivnuyu sostavlyayushchie
(aktivnogo i passivnogo agenta) harakteristiki ustanovivshegosya
avtokolebatelp1nogo interaktivnogo protsessa podrazdelyayut\-sya na
graficheskie dannye o dinamicheskom sostoyanii passivnogo agenta i
chastotno-amp\-li\-tud\-nye harakteristiki aktivnogo agenta.

Virtualizatsiya estestvennoe0 interaktivnoe0 psihoinformatsionnoe0\linebreak
sistemy takim obrazom zaklyuchaet\-sya v zadanii algoritma vtorichnogo sinteza
[3] graficheskih dannyh o dinamicheskom sostoyanii passivnogo agenta pri
lyubom interaktivnom protsesse po zadannoe0 sovokupnosti uka\-zan\-nyh dannyh
pri
ustanovivshihsya avtokolebatelp1nyh protsessah iz ne\-kotorogo mnozhestva, a
takzhe so\-ot\-vet\-st\-vu\-yu\-shchih chastotno-amplitudnyh ha\-rakteristik
aktivnogo agenta v e1tih protsessah.

Tem samym printsipialp1naya s\-hema virtualizatsii estestvennyh
in\-te\-rak\-tiv\-nyh psihoinformatsionnyh sistem, osnovannoe0 na
ispolp1zovanii\linebreak vtorichnogo sinteza izobrazhenie0, imeet vid:

$$\boxed{\aligned&\boxed{{\aligned &\foldedtext\foldedwidth{0.8in}{\cyrf
Estestvennaya
\newline  interaktivnaya\newline psihoinforma-\newline tsionnaya sistema}\\
&\boxed{\foldedtext\foldedwidth{0.6in}{\cyrf Passivinye0\newline agent
(obp2ekt)}}\\
&\boxed{\foldedtext\foldedwidth{0.6in}{\cyrf Aktivnye0\newline agent
(subp2ekt)}}\endaligned}\
\boxed{\aligned &\foldedtext\foldedwidth{1in}{\cyrf Interaktivnye0\newline
protsess v\newline avtokolebatelp1nom\newline rezhime}\\&\to\boxed{\foldedtext
\foldedwidth{0.63in}{\cyrf Graficheskie\newline
dannye}}\to\\&\to\boxed{\foldedtext\foldedwidth{0.63in}{\cyrf
Chastotno-\newline amplitudnye\newline harakteristiki}}\nearrow\endaligned}}
\aligned&\\&\Rightarrow
\boxed{\foldedtext\foldedwidth{0.45in}{\cyrf Algoritm\newline
vtorichnogo\newline
sinteza}}
\Rightarrow\endaligned
\boxed{\aligned &\foldedtext\foldedwidth{1.4in}{\cyrf Iskusstvennaya
interaktivnaya\newline kompp1yuterograficheskaya\newline psihoinformatsionnaya
\newline sistema}\\&\boxed{\foldedtext\foldedwidth{1.3in}{\cyrf Dinamicheskoe
izobrazhenie\newline proizvolp1nogo$^{*)}$ interaktivnogo\newline protsessa v
intentsionalp1noe0\newline anomalp1noe0 virtualp1noe0
\newline realp1nosti}}\endaligned}\\&
\\& ^{*)} \text{\cyrs Ne obyazatelp1no
avtokolebatelp1nogo}\endaligned}$$
\pagebreak

\definition{\cyb (Kriterie0 adekvatnosti izobrazheniya i korrektnosti
vir\-tu\-a\-li\-za\-tsii)} \cyr Virtualizatsiya estestvennoe0 interaktivnoe0
psihoinformatsionnoe0 sistemy yav\-lya\-et\-sya korrektnoe0, esli
poluchayushcheesya v rezulp1tate nei0 izobrazhenie e1toe0 sistemy posredstvom
nekotoroe0 in\-te\-rak\-tiv\-noe0 psihoinformatsionnoe0 sistemy adekvatno
is\-hodnoe0 sisteme, to estp1 pri odnovremennom funktsionirovanii kak
iskusstvennoe0 tak i estestvennoe0 in\-te\-rak\-tiv\-nyh sistem s odnim i tem
zhe aktivnym agentom vosproizvedenie izobrazheniya zadannogo
in\-te\-rak\-tiv\-no\-go protsessa pozvolyaet vyzvatp1 ego protekanie v
estestvennoe0 interaktivnoe0 psihoinformatsionnoe0 sisteme.
\enddefinition

Otmetim, chto my vopros o tom, kak vozmozhna korrektnaya
vir\-tu\-a\-li\-za\-tsiya, ne stavit\-sya v dannoe0 rabote; korrektnostp1
kazhdoe0 vir\-tu\-a\-li\-za\-tsii i adekvatnostp1 kazhdogo izobrazheniya mogut
bytp1
provereny lishp1 e1ksperimentalp1no.

\Refs\nofrills {\cyb Spisok literatury}
\roster
\item"[1]" \cyre Yurp1ev D.V., Kompleksnaya proektivnaya geometriya i
kvantovaya proektivnaya teoriya polya. Teor. matem. fizika. 1994. T.00.
S.00-00.
\item"[2]" \rm Juriev D., Octonions and binocular mobilevision; {\it
hep-th/9401047};\newline
Juriev D., Visualizing 2D quantum field theory: geometry and informatics of
mobilevision; {\it hep-th/9401067}.
\item"[3]" \cyre Yurp1ev D.V., Vtorichnye0 sintez izobrazhenie0 v e1lektronnoe0
kompp1yuternoe0 fotografii; {\it adap-org/9409002}.
\endroster
\endRefs
\enddocument